\documentclass[reqno]{amsart}
\usepackage{amsmath,amsfonts,amssymb,amscd}
\usepackage{mathrsfs}
\usepackage[numbers]{natbib}
\usepackage{bm,slashed}

\theoremstyle{plain}
\newtheorem{theorem}{Theorem}
\newtheorem{lemma}[theorem]{Lemma}

\theoremstyle{definition}
\newtheorem{example}{Example}

\title[Reflection Positivity and Monotonicity]{Reflection Positivity and Monotonicity}

\author{Arthur Jaffe and Gordon Ritter}
\address{Harvard University\\
         17 Oxford St., Cambridge, MA 02138}
\email{Arthur\_Jaffe@harvard.edu, ritter@post.harvard.edu}

\date{May 3, 2007}

\newcommand{\cN}{\mathscr{ N}}

\newcommand{\sC}{\mathscr{ C}}

\newcommand{\sH}{\mathscr{ H}}

\newcommand{\sO}{\mathscr{ O}}

\newcommand{\sS}{\mathscr{ S}}

\newcommand{\C}{\mathbb{C}}

\def\R{\mathbb{R}}

\def\eps{\epsilon}

\def\th{\theta}
\def\vth{\vartheta}
\def\ga{\gamma}

\newcommand{\Om}{\Omega}
\def\al{\alpha}

\newcommand{\Ga}{\Gamma}
\def\de{\delta}
\def\De{\Delta}
\newcommand{\om}{\omega}

\newcommand\si{\sigma}

\newcommand{\lrp}[1]{\left( #1 \right)}
\newcommand{\lrpbig}[1]{\big( #1 \big)}
\newcommand{\lrpBig}[1]{\Big( #1 \Big)}

\newcommand{\lrb}[1]{\left[ #1 \right]}

\newcommand{\lrabig}[1]{\big\langle #1 \big\rangle}
\newcommand{\lraBig}[1]{\Big\langle #1 \Big\rangle}

\newcommand{\absBig}[1]{\Big| #1 \Big|}

\renewcommand{\d}{\partial}

\def\<{\langle}
\def\>{\rangle}
\newcommand\del{\nabla}

\newcommand\To{\longrightarrow}

\newcommand\supp{\operatorname{supp}}

\newcommand{\diag}{\operatorname{diag}}

\newcommand{\ds}{\displaystyle}

\newfont{\boldit}{cmbxti10}

\newcommand\te{\text}

\newcommand\f{\frac}
\newcommand\ol{\overline}

\newcommand\bea{\begin{eqnarray}}
\newcommand\eea{\end{eqnarray}}
\newcommand\beas{\begin{eqnarray*}}
\newcommand\eeas{\end{eqnarray*}}

\newcommand{\er}[1]{\eqref{#1}}
\newcommand{\be}{\begin{equation}}
\newcommand{\bel}[1]{\begin{equation}\label{#1}}
\newcommand\ee{\end{equation}}

\newcommand\nn{\nonumber}

\def\dsl{{\slashed{\partial}}}
\def\psl{{\slashed{p}}}

\renewcommand{\div}{\operatorname{div}}

\newcommand{\Cl}{\operatorname{Cl}}
\renewcommand{\div}{\operatorname{div}}

\newcommand{\mat}[4]{\left(\begin{smallmatrix}#1 & #2 \\ #3 & #4\end{smallmatrix}\right)}

\newcommand{\vpsqr}{{{\vec p}{}^{\ 2}}}
\newcommand{\fm}{{\mathfrak{m}}}

\newcommand{\dV}{\operatorname{dV}}
\newcommand{\dS}{\operatorname{dS}}
\newcommand{\vol}{\mathrm{vol}}

\begin{document}

\begin{abstract}
We prove general reflection positivity results for both scalar
fields and Dirac fields on a Riemannian manifold, and comment on
applications to quantum field theory. As another application, we
prove the inequality $C_D \leq C_N$ between Dirichlet and Neumann
covariance operators on a manifold with a reflection.
\end{abstract}

\maketitle

\section{Introduction}
Reflection positivity (RP) provides the fundamental relation
between functional integration and quantization.  Osterwalder and
Schrader formulated this notion in an attempt to understand the
special case discovered by Symanzik
\cite{Symanzik:jmp}---elaborated by Nelson
\cite{Nelson:1973a,Nelson:1973b}, and by many authors
since---between Markov random fields and quantum fields. The
Osterwalder-Schrader theory not only pertains to classical
probability theory, but also makes it possible to incorporate
theories with spin (fermions and gauge theory), and provides the
possibility to quantize differential forms. In quantum theory it
leads to an analysis of the Hamiltonian and to other symmetry
groups, see for instance
\cite{Fr:80,KL:81,JR:Symmetries,Dimock:2003an}. RP also pertains
to the framework of statistical physics on a lattice, where it
leads to an analysis of the transfer matrix. As a result of the
importance of RP, several different ways to understand it appear
in the literature.

In this paper we analyze some properties of RP, monotonicity, and static
space-times.  In particular we analyze RP arising from the Green's function for
the Laplace operator under general conditions, leading to a positive inner
product distinct from the standard positive inner product given by the Green's
function for the Laplacian.

We also analyze the case of a general Dirac operator compatible with time
reflections.  This case presents new phenomena, as the Green's function for the
Euclidean Dirac operator is not positive. However, using time-reflection, we
establish the existence of a positive inner product and a corresponding Hilbert
space, providing a general framework for quantization in this case as well.

\section{Reflections and the Laplace Operator}
\label{Sec:GeneralRP}

Let $M$ be a complete, connected Riemannian manifold without
boundary, and  with isometry group $G$. Let $U$ denote the natural
unitary representation of $G$ on $L^2(M)$, defined on a dense
domain by
\bel{defu}
    U_\psi f \ :=\  f^\psi = f \circ \psi^{-1}
    \ \ \te{ for } \ \ \psi \in G .
\ee

Let $\del$ denote the Levi-civita connection on $M$ associated to
the metric. Let $\De = \De_M = \del^*\del$ denote the
(negative-definite) covariant Laplacian, defined initially on the
domain $C_c^\infty(M)$ of smooth functions of compact support.
Under our assumptions, it follows that $\De$ is essentially
self-adjoint on this domain (\cite{Gaffney:1951}; see also
\cite{chavel:book}), and so naturally we consider the unique
self-adjoint extension and use the spectral theorem accordingly.
The operator $U_\psi$ commutes with $\De$, which can be seen by
writing $\De = d^* d$ on 0-forms. The resolvent $C = (-\De +
m^2)^{-1}$ is a bounded operator on $L^2(M)$. It also
follows\footnote{See for instance \cite[Theorem III.6.5]{Kato}.}
that $[U_\psi, C] = 0$, which becomes $C f^\psi = (Cf)^\psi$ in
the notation of eq.~\er{defu}.

For $\th \in G$, the \emph{fixed-point set} is the set
\[
    M^\th  = \{ p\in M : \th(p) = p \}.
\]
The isometry $\th$ is said to be a \emph{reflection} if $d\th_p$
is a hyperplane reflection in the tangent space for some $p \in
M^\th$. In this case, $M^\th$ is a disjoint union of totally
geodesic submanifolds, at least one of which is of codimension one
\cite{MichorReflections}. Any codimension-one component of $M^\th$
is called a \emph{reflection hypersurface}.

To formulate a general notion of reflection positivity, let $M$ be
a complete connected Riemannian manifold with a reflection $\th$.
Let $\Om \subset M$ be a submanifold with boundary $\d\Om$, such
that $\d\Om$ is contained in a union of reflection hypersurfaces.
Let $f \in L^2(M)$ be a complex-valued function with support
$\sS_f$, which is of class $C^2$ (i.e. all second derivatives are
continuous), such that
\bel{support-condition}
    \th(\sS_f) \subseteq \Om
    \ \ \te{ and } \ \
    \vol(\Om \cap \sS_f) = 0.
\ee
We do {\bf not} assume that $\th$ is \emph{disecting}, i.e. that
$M \setminus M^\th$ is disconnected.

\begin{example}
Choose coordinates $(t, \vec x)$ on $\R^d$, and define $\R^d_+$
and $\R^d_-$ to be the half-spaces with $t \geq 0$ or $t \leq 0$
respectively. Let $f$ be such that $\sS_f \subseteq \R^d_+$, and
$\Om \subseteq \R^d_-$ with $\d\Om \subseteq \{ t = 0 \}$.
\end{example}

\begin{example}
Let $T$ be a Riemann surface with an antiholomorphic involution
$\th : T \to T$, such as $\th (z) = 1/\ol{z}$ on the Riemann
sphere. Accordingly, write $T = T_- \cup T_+$, where $T_{\pm}$ are
closed, $T^\th = \d T_\pm$, and $\th : T_+ \to T_-$. Let $\sS_f
\subseteq T_+$ and $\Om \subseteq T_-$ with $\d\Om \subset T^\th$.
For the Riemann sphere, $T^\th$ is the unit circle $|z|=1$.
\end{example}

\begin{theorem} \label{theorem:GeneralRP}
Let $M$ be complete and connected with a reflection $\th$. Let $f$
and $\Om$ be as above. Then
\bel{eqn:GeneralRP}
    \boxed{
   0\le  \<f^\th, Cf\>  \, .
    }
\ee
\end{theorem}

\begin{proof}
For $u,v : M \to \C$, let $(u,v) = \ol{u} \, v \, \dV$, where
$\dV$ denotes the natural Riemannian volume measure on $M$.
%Our assumptions imply that
%$\supp(f^\th) = \th(\Om_f) \subseteq \Om$.
Define $u = Cf$ and note that, by eq.~\er{support-condition},
$C^{-1} u^\th = (C^{-1}u)^\th = f^\th$ has support in $\Om$. Hence
\bel{Forms1}
    \< u, f^\th \> = \int_\Om (u, C^{-1} u^\th)
    =
    \int_\Om [(u, C^{-1} u^\th) - (C^{-1} u, u^\th)],
\ee
where the second equality holds because $C^{-1}u = f$ is zero a.e.
in $\Om$. Let $n$ denote the unit normal vector to $\d\Om$.
%The
%standard integration-by-parts formula for the covariant derivative
%is:
%\[
%    \int_\Om (u \De v - v \De u)\, \dV
%    =
%    \int_{\d\Om} \lrpbig{u \del_n v - v \del_n u}\, \dS \,.
%\]
Now in \er{Forms1}, replace $C^{-1}$ with $-\De + m^2$ and
integrate by parts to find
\bel{Forms2}
    \< f^\th, u \>
    =
    \int_{\Om} [(\De u, u^\th) - (u, \De u^\th)]
    =
    \int_{\d\Om} \big[ u^\th \del_n \ol{u} - \ol{u}\, \del_n u^\th \big] \dS .
\ee

For $p \in \d\Om$, it clear that $d\th_p = d\th^{-1}_p =
\diag(-1,1,\ldots, 1)$ in a coordinate basis for $T_p M$ where the
first coordinate is in the direction of $n_p$ and the other
directions are tangential to $\d\Om$. Hence $(\del_n u^\th)_p = -
(\del_n u)_p$. Using this (and the identity $u^\th = u$ on
$\d\Om$) to simplify the second term in \er{Forms2}, we have
\[
    \< f^\th, Cf \>
    =
    2 \Re \int_{\d\Om} u\, \del_n \ol{u}\ \dS
\]
where $\Re$ denotes the real part. We now show that this quantity
is real (and positive), completing the proof.
\begin{eqnarray}
    \int_{\d\Om} (u\, n_a \del^a \ol{u})\ \dS
    &=&
    \int_{\d\Om} (u\, n_a \del^a \ol{u})\ \dS
    =
    \int_{\Om} \del_a (u\, \del^a \ol{u})\ \dV
    \nonumber\\
    &=&
    \int_{\Om} \lrp{ \del_a u \,\del^a \ol{u} + u \Delta \ol{u}}\ \dV
    \nonumber\\
    &=&
    \int_{\Om} \lrp{|\del u|^2 + m^2 |u|^2}\  \dV \geq 0 .
    \quad\phantom{1}
    \label{EuclideanAction}
\end{eqnarray}
To obtain \er{EuclideanAction} we used that $\De u = m^2 u$ a.e.
on $\Om$, which holds since $\Om \cap \sS_f$ has measure zero.
\end{proof}

Theorem \ref{theorem:GeneralRP} has applications to quantum field
theory. For curved spacetimes which possess both a Riemannian and
a Lorentzian section (such as the Schwarzschild black hole),
eq.~\er{eqn:GeneralRP} is the inner product in the one-particle
space for scalar fields, and the positivity of this inner product
is one of the cornerstones of the Euclidean approach. This was
discovered by Osterwalder and Schrader \cite{OS1,OS2} for $\R^d$,
and generalized to curved spacetimes in
\cite{JR:CurvedSp,JR:Symmetries}. From the proof of Theorem
\ref{theorem:GeneralRP}, we see that $\<f^\th, Cf\>$ is twice the
Euclidean action applied to the potential $u = Cf$, in the region
$\Om$.

The action functional for a general scalar quantum field theory on
a curved background may include a term of the form $\xi R$, where
$R$ is the Ricci scalar, and $\xi$ is a real coupling constant.
The special case $\xi = 0$ is called \emph{minimal coupling}; we
now discuss the general case.

Let $M$ be a complete connected  Riemannian manifold with a
reflection $\th$. Let $R$ be the Ricci scalar and $\xi \in \R$ be
such that
\bel{CurvatureBound}
    0< m^2 + \xi R\;,
\ee
everywhere on $M$. Then $-\De + m^2 +\xi R$ has a self-adjoint
extension which is invertible, and thus we define
\[
    C_\xi = (-\De + m^2 +\xi R)^{-1} .
\]

\begin{theorem}
Let $M$ be a complete connected Riemannian manifold with a
reflection $\th$, and assume the curvature bound
\er{CurvatureBound}. Let $f$ be as above. Then
\bel{Hamiltonian}
   0\le \< f^\th, C_\xi f\> \; .
\ee
\end{theorem}

\begin{proof}
Following the same steps as leading to  \er{EuclideanAction}, we
have
\[
    \< f^\th, C_\xi f\>
    =
    2\int_\Om \lrb{ |\del u|^2 + (m^2 +\xi R)|u|^2 } \, \dV .
\]
The conclusion follows.
\end{proof}

\section{Comparison of Dirichlet and Neumann Covariance}
\label{sec:Comparison}

Glimm and Jaffe \cite{GJ,NoteOnRP} discovered that reflection
positivity for free Euclidean fields is equivalent to the
operator-monotonicity  of the Green's operator $C$, as one varies
boundary conditions on the $t=0$ plane. More precisely $C_D\le
C_N$, where $D,N$ denote respectively the classical Dirichlet or
Neumann boundary conditions at $t=0$. The proof remarks that
Green's functions satisfying Dirichlet and Neumann boundary
conditions can be obtained using mirror charges, and these
representations lead to reflection positivity. De Angelis, de
Falco, and Di Genova \cite{DDD86} used this property to give a
simple proof of reflection positivity for manifolds with an
isometric involution $\th$; we also use this method here.

We first discuss Dirichlet and Neumann conditions on manifolds in
general, and then in the special case of a reflection, give a
simple proof of the fundamental inequality $C_D \leq C_N$.

\begin{lemma} \label{lem:DNRepresentation}
Let $M$ be complete and connected with a reflection $\th$. Let
$\sO \subset M$ be a submanifold with boundary $\d\sO \subseteq
M^\th$. Let $C_{D,N}$ denote the resolvent of the Laplace operator
on $L^2(\sO)$ with either Dirichlet (D) or Neumann (N) boundary
conditions on $\d\sO$. Then
\[
    C_D = (I-U_\th)C, \quad C_N = (I+U_\th)C, \ \ \te{ and } \ \
    U_\th C = \f12 (C_D - C_N) \ \ \te{on} \ \ L^2(\sO) .
\]
\end{lemma}

\begin{proof}
Write $C$ in integral form with kernel $\sC$, so that
\[
    (Cf)(x) = \int_M \sC(x, y) f(y)\ d\vol_y
\]
where $\vol$ is the Riemannian volume measure; in coordinates
$d\vol_x = (\det g_{ab})^{1/2} dx$. Note that $\sC : M \times M
\to \R$ is not defined on the diagonal $x = y$. The two
fundamental properties of the kernel $\sC$ are invariance under
the diagonal action of $G \subset G \times G$, and that it is a
Green's function. Thus
\[
    \sC(g x, g y) = \sC(x,y)
    \ \ \forall\, g \in G,
    \quad \te{ and } \quad
    (-\De_x + m^2)\sC(x,y) = \delta_x(y) \, .
\]
To prove the second property, write
\[
    f(x) = ((-\De + m^2)Cf)(x) = \int (-\De_x + m^2)\sC(x,y) f(y)
    \ d\vol_y.
\]
Then by definition, $(-\De_x + m^2)\sC(x,y) = \delta_x(y)$ as
distributions.

Since $[C,U_{\th}]=0$, the integral kernel of $U_{\th} C$ is
$\sC(\th x, y)$. Thus the kernel of $(I - U_{\th})C$ is
\[
    k_-(x,y) := \sC(x,y) - \sC(\th x, y) \, .
\]
Clearly for $x\in \d\sO$ or $y \in \d\sO$, we have $k_-(x,y) = 0$,
so $k_-$ satisfies Dirichlet boundary conditions. Also,
\[
    (-\De_x + m^2) k_-(x,y) = \delta_x(y) - \delta_x(\th y).
\]
For $x \in \sO$, it follows that $\de_x \circ \th$ vanishes for
test functions supported in $\sO$, and hence $k_-$ is the
Dirichlet Green's function in $\sO$. Now
\[
    k_+(x,y) := \sC(x,y) + \sC(\th x, y)
\]
is also a Green's function in $\sO$ for the same reason. Let
$\sC_y(x) = \sC(x,y)$ and let $\d_n$ denote the normal derivative
in the variable $x$ on the boundary $\d\sO$. Then
\[
    \d_n(\sC_y^\th)|_p = - \d_n \sC_y|_p \ \te{ for } \ p \in
    \d\sO.
\]
It follows that $\d_n k_+ = 0$ on $\d\sO$, so $k_+$ is the Neumann
Green's function on $\sO$, completing the proof.
\end{proof}

We now prove the operator inequality stated previously. The result
is known for $M = \R^d$, though the proof which has appeared in
the literature is complicated due to delicate issues concerning
the domains of self-adjoint operators and associated quadratic
forms. We present a simpler proof that also generalizes to
manifolds.

\begin{theorem}\label{thm:monotonicity}
Let $M$ be complete and connected with a reflection $\th$. Let
$\sO$ be a submanifold with boundary $\d\sO \subseteq M^\th$. Then
\[
    C_D \leq C_N \ \ \te{ on } \ \  L^2(\sO).
\]
\end{theorem}

\begin{proof}
By Lemma \ref{lem:DNRepresentation}, for $f \in C_c^\infty(\sO)$,
we have
\bel{cncd}
     \< f, (C_N - C_D)f\> = 2 \< f, U_\th Cf\> .
\ee
Now apply Theorem \ref{theorem:GeneralRP} with $\Om = \sO^c \cup
\d\sO$. The boundary of $\Om$ is the same as the boundary of
$\sO$, and the common boundary is contained in $M^\th$. The
support $\sS_f$ of $f$ is disjoint from $\Om$ (up to sets of
measure zero), so Theorem \ref{theorem:GeneralRP} can be applied.
Thus \er{cncd} is positive and $C_D \leq C_N$ as desired.
\end{proof}

\section{The Dirac Operator}

A certain sense of mystery surrounds Euclidean fermions.  It
revolves about two issues, the more elementary of which is whether
the Euclidean Green's functions are reflection positive. In the
case of Pfaffian or determinantal imaginary-time Green's
functions, this reduces to the question of reflection positivity
for the pair correlation function. In the following, we resolve
this question in the affirmative, giving a proof that is at once
very simple, and very general; our proof applies to any bundle of
Clifford modules over a static manifold.

\subsection{Clifford Bundles}
\label{sec:diracop-mfld}

Let $M$ be a Riemannian manifold. The Clifford algebra of the
cotangent space $T_x^* M$ (with its natural inner product) will be
denoted $\Cl_x$, and the association of the vector space $\Cl_x$
to the point $x$ defines the Clifford bundle $\Cl(M) \to M$.

Now suppose $E \to M$ is a Hermitian vector bundle such that each
fiber $E_x$ is a Hermitian $\Cl_x$-module in a smooth fashion. Let
$\Ga(E)$ denote the space of smooth sections, and let $\Ga(E;
\sO)$ denote the space of local sections over an open set $\sO
\subset M$. Extend the notation to allow $\sO$ to be a submanifold
with boundary.

Let $E \to M$ be endowed with a connection $\del$. Since $E_x$ is
a $\Cl_x$-module, the inclusion $T_x^*(M) \subset \Cl_x$ gives
rise to a natural bundle map $\fm : T^* \otimes E \to E$ called
\emph{Clifford multiplication}. Explicitly, we have a sequence
\bel{CliffMult}
    \begin{CD}
    \Ga(E)
    @>\del>>
    \Ga(T^* \otimes E)
    @>\fm>>
    \Ga(E).
    \end{CD}
\ee
Denote Clifford multiplication simply by $\xi \cdot v := \fm(\xi
\otimes v)$. Composing the maps \er{CliffMult} gives the
\emph{Dirac operator}
\[
    \dsl \ =\
    \fm\del\  :\
   \Ga(E) \To \Ga(E).
\]

Many computations are facilitated by the use of local coordinates.
Let $\sO$ be an open subset of $M$ on which we have defined an
orthonormal frame $\{ e_j \}$ of tangent vector fields, and let
$\{ v_j \}$ denote a dual coframe of 1-forms. For $\phi \in
\Ga(E;\sO)$, the above definitions imply that
\[
    \dsl\phi = \sum_j v_j \cdot \del_{e_j} \phi .
\]

A \emph{Clifford connection} on $E$ is a metric connection $\del$
that is a derivation with respect to Clifford multiplication, i.e.
\[
    \del_X(v \cdot s) = (\del_X v) \cdot s + v \cdot \del_X s
\]
for a vector field $X$, a 1-form $v$, and a section $s$. In the
first term, $\del_X v$ denotes the Levi-civita connection on $M$,
while in the second term $\del_X$ denotes the connection on $E$.

If $\del$ is a Clifford connection on a boundaryless manifold,
then $\dsl$ is a skew-symmetric operator on the domain of smooth,
compactly supported sections \cite[Prop.~1.1, p.~246]{Taylor}.

\subsection{Reflection Positivity}

Let $M$ be a complete Riemannian manifold. Further assume $M$ is
\emph{static}; then there are coordinates $(x^i \, :\, 0 \leq i
\leq d-1)$ such that $\d / \d x^0$ is a hypersurface-orthogonal
Killing field. In many examples from physics, $x^0$ plays the role
of (Euclidean) time, so we also write $t = x^0$.

Corresponding to the local frame $\d / \d x^i$ is a dual frame of
one-form fields, $dx^i$. Let $\gamma^i$ denote the operator of
Clifford multiplication by $dx^i$, so that $\ga^i(v) = dx_i \cdot
v$. Then
\bel{CliffManifold}
    \{ \ga^i, \ga^j\} = 2 g^{ij} I\;,
\ee
where $g^{ij}$ is the inverse metric, and $I$ is the identity on
fibers of $E$. Since the coordinate $t = x^0$ is determined (up to
a constant) by the geometry, the operator $\ga^0$ does have a
coordinate-free meaning, whereas in general, $\ga^i$ for $i \ne 0$
are coordinate-dependent.

Locally, a static metric takes the form
\bel{staticmetricFG}
    ds^2 = F(x) dt^2 + G_{ab}(x) dx^a dx^b .
\ee
where $F$ and $G_{ab}$ are $t$-independent functions. After an
arbitrary choice of a time-zero slice $\Sigma = \{ t = 0\}$, $M$
has the structure
\[
    M = \Om_- \cup \Sigma \cup \Om_+,
    \qquad \d\Om_\pm = \Sigma \;.
\]
Let $\eps : \Om_+ \to \Om_-$ be the natural reflection map, which
in coordinates is given by
\[
    \eps(t, y) = (-t,y),
\]
where $y$ is a coordinate on $\Sigma$. This induces a pullback map
$\eps^*$ acting on sections of $E$. Let $\vth = \ga^0 \eps^*$.
Note that
\bel{anticom}
    \{ \vth, \dsl \} = \big\{ \ga^0 \eps^*, \sum_j \ga^j \del_{e_j} \big\}
    =
    0\;.
\ee
To prove \er{anticom}, note that the $j=0$ term vanishes since $\{
\eps^*, \del_{e_0} \} = 0$, while the other terms vanish because
$\{\ga^0,\ga^j\} = g^{0j} = 0$.

\begin{theorem}\label{thm:RP-Dirac-Curved}
Let $E \to M$ be a holomorphic Clifford bundle with Clifford
connection $\del$. For a smooth section $\phi \in \Ga(E; \Om_+)$
supported in $\Om_+$, we have
\[
    0 \le \< \vth \phi, (\dsl - m)^{-1}\phi \> \;.
\]
\end{theorem}

\begin{proof}
Let $u=(\dsl - m)^{-1}\phi$ and $u^\vth = \vth u$, so we have
\begin{eqnarray}
    \< \vth \phi, (\dsl - m)^{-1}\phi \> &=& \< \vth (\dsl - m)u, u\>
    =
    \< \vth \dsl u, u\> - m \< u^\vth,u \>
    \nonumber\\
    &=&
    -\int_{\Om_-} [\< \dsl u^\vth, u\> + m \< u^\vth,u \>]
    \nonumber\\
    &=&
    -\int_{\Om_-} [\< \dsl u^\vth, u\> + m\ \< u^\vth,u \> +
    \< u^\vth,(\dsl-m)u\>] \qquad\qquad
    \label{extraterm1}\\
    &=&
    -i\int_{\Om_-} [\< D u^\vth, u\> - \< u^\vth, Du\> ]
    \label{Expr2}
\end{eqnarray}
where we used $\{ \vth, \dsl \}=0$, and to obtain \er{extraterm1}
from the previous line, we used that $(\dsl-m)u = 0$ on $\Om_-$.

By \cite[p.~247]{Taylor}, for any sections $\al,\beta$ of $E$ we
have
\[
    \div X = -i[\< D\al, \beta\> - \< \al, D\beta\>]
\]
where $X$ is the vector field defined by $\<X,v\> = \<\al,
v\cdot\beta\>_E$ for $v \in \Om^1(M)$. Apply this with $\al =
u^\vth$ and $\beta = u$ so, using \er{Expr2}, we have
\[
    \< \vth \phi, (\dsl-m)^{-1}\phi \>
    =
    \int_{\Om_-} \div X \ \dV ,
\]
where $\<X,v\> = \<\vth u, v\cdot u\>_E$ and $\dV$ is the volume
element on $M$.

On $\Sigma$, the outward-pointing unit normal to $\Om_-$ is the
Killing vector $\d_t$ divided by its norm, i.e. $\hat n = F^{-1/2}
\d_t$, where $F$ is defined in eq.~\er{staticmetricFG}. Let $\nu$
denote the 1-form dual to $\hat n$, so $\nu = \sqrt{F} dx^0$. Then
by the divergence theorem,
\[
    \int_{\Om_-} \div X \ \dV
    =
    \int_{\d \Om_-} \< X,\nu\>\ \dS .
\]
On $\Sigma = \d\Om_-$, $\eps$ is the identity map and so $u^\vth =
\ga^0(u)$. Also, $\nu\cdot s = \sqrt{F}\, \ga^0(s)$, so we have
\[
    \lrabig{ X,\, \nu}
    =
    \lrabig{\vth u,\, \nu \cdot u}_E
    =
    \lrabig{\ga^0(u),\, \sqrt{F} \ga^0(u)},
    \ \
    (\te{on } \Sigma).
\]
Hence
\bel{final-positivity}
    \lrabig{ \vth \phi,\, (\dsl-m)^{-1}\phi }
    =
    \int_{\Om_-} \div X \ \dV
    =
    \int_{\d\Om_-} \lrabig{\ga^0(u),\, \sqrt{F} \ga^0(u)} \ \dS
    \geq 0.
\ee
\end{proof}

The power of Theorem \ref{thm:RP-Dirac-Curved} lies in its
generality; the result is valid for \emph{any Clifford connection
on any vector bundle} over a static manifold. This includes as
particular examples the Dirac operator on the spinor bundle
$S(\tilde P)$ over a manifold with a spin structure $\tilde P \to
M$, as well as the ``twisted Dirac operator'' $D_F$ on the tensor
product $E = S(\tilde P)\otimes F$, where $F$ is a bundle with
metric connection.

As a corollary to Theorem \ref{thm:RP-Dirac-Curved}, we infer the
existence of a Hilbert space $\sH_{D}$ whose inner product is
given by
\[
    (s,s')_D = \< \vth s, (\dsl - m)^{-1} s' \>.
\]
Precisely, $\sH_{D}$ is the completion of the coset space $\Ga(E;
\Om_+) / \cN_D$, where $\cN_D$ is the kernel of the form $(\cdot ,
\cdot)_D$. The space $\sH_{D}$ can be interpreted as the
one-particle space for a theory of fermions on the spacetime $M$.

%--------------------------------------------------------------------

\subsection{Flat Spacetimes}

It is very useful to see the abstract framework of the last two
sections worked out in the explicit example of $M = \R^d$. In this
case, we also prove reflection positivity by Fourier analysis.

For an integral operator $C$ on $L^2(\R^d)$, we use the convention
\[
    Cf(x) = \int C(x,y)f(y) dy.
\]
For $C = (-\De + m^2)^{-1}$, the kernel is translation-invariant,
so we write $C(x,y) = C(x-y)$, and we may obtain the latter
explicitly via the Fourier transform
\[
    C(x,y) = C(x-y)
    =
    (2\pi)^{-d} \int \f{e^{-i p (x-y)}}{p^2+m^2} \, dp.
\]
Note that $C(x-y) = \ol{C(y-x)} = C(y-x)$. It follows that the
integral kernel of $\ds\d_x C$ is equal to $\ds\d_x C(x-y)$.

Let $\ga_j$, for $j = 0, \ldots, d-1$, be a set of Hermitian
operators\footnote{An example in $d=4$ is $\ga_0 =
\mat{0}{I}{I}{0},$ and $\ga_j = i \mat{0}{\si_j}{-\si_j}{0}$ for
$j=1,2,3$.} on a complex Hilbert space $V$ satisfying:
\[
    \{ \ga_i, \ga_j \} = 2 \de_{ij}\, I.
\]
Denote $\psl = \sum_{i=0}^{d-1} \ga_i p_i$, with $\dsl$ defined
similarly. This arises from the general theory of Section
\ref{sec:diracop-mfld}, by setting $E = \R^d \times V$, a trivial
Hermitian vector bundle over $\R^d$ with standard Riemannian and
Clifford structures. Then
\[
    \psl^2 = \f12 \{ \psl, \psl \}
    =
    \f12 \sum_{a,b} p_a p_b \{ \ga_a, \ga_b \} = p^2 I.
\]
Similarly, $\dsl^2 = \De$, so $-(\dsl + m)(\dsl - m) = -\De+m^2 =
C^{-1}$ and hence
\[
    (\dsl - m)^{-1} = -(\dsl + m )C.
\]
Let $\eps(x_0, \vec x) = (-x_0, \vec x)$ be a coordinate
reflection. For $f : \R^d \to V$, define
\bel{def-vth}
    (\vth f)(x) \ :=\  \ga^0 f(\eps x) .
\ee
It follows that $\vth$ is a self-adjoint operator on $L^2(\R^d,V)$
with $\vth^2 = I$.

\begin{theorem}\label{RP-flatspace}
The operator $(\dsl-m)^{-1}$ is reflection-positive in the sense
that
\bel{RPDirac}
    \lrabig{ \vth f, (\dsl-m)^{-1} f } \geq 0
    \ \
    \te{ for }
    \ \
    \supp f \subseteq \{ x_0 > 0 \} .
\ee
\end{theorem}

We give two proofs of Theorem \ref{RP-flatspace}; one by Fourier
analysis and one by potential theory.

\begin{proof}[Proof (Fourier analysis)]
By direct calculation,
\[
    -\ga^0 (\dsl + m) C(x) = -\ga^0 \int dp_0 d\vec p \,
    \lrpBig{ \ga^0  (-i p_0) + \sum_{j>0} \ga^j (-i p_j) + m }
    \f{e^{-i px}}{p^2 + m^2} \, ,
\]
By contour integration, for any $t \in \R$ we have:
\[
    \int \f{e^{-i p_0 t}}{p_0^2 + \omega^2} \ dp_0
    =
    \f{\pi e^{-|t| \om}}{\om},
    \qquad
    \int p_0 \, \f{e^{-i p_0 t}}{p_0^2 + \omega^2} \ dp_0
    =
    -i \pi e^{-|t| \om} .
\]
We use these formulas with $\om = (\vpsqr + m^2)^{1/2}$, and $t =
x_0$. So,
\begin{eqnarray*}
    -\ga^0 (\dsl + m) C(x)
    &=&
    \pi \int
    \lrpBig{ e^{-|t| \om} + \f{e^{-|t| \om}}{\om} \Big[
    \sum_{j=1}^{d-1} i \ga_0 \ga^j p_j -m \ga_0 \Big] }
    e^{-i \, \vec p \cdot \vec x} \, d\vec p
    \nonumber \\
    &=&
    \pi \int
    \f{e^{-|t| \om}}{\om} A\, e^{-i \, \vec p \cdot \vec x}\, d\vec p ,
\end{eqnarray*}
where we define
\[
    \vec\eta \ :=\  i \ga^0 \vec\ga,
    \ \
    \om = (\vpsqr + m^2)^{1/2},
    \ \ \te{ and } \ \
    A = \omega I + \vec\eta \cdot \vec p -m \ga_0
    \, .
\]
Here, each component $\eta_j$ and $A$ are $d \times d$ matrices,
and $A$ has $\vec p$-dependent matrix elements. The matrix $\Om =
\vec\eta \cdot \vec p -m \ga_0$ is Hermitian with $\Om^2 = \om^2
I$, hence $A = \omega I + \Om$ has eigenvalues $0, 2\om$. In
particular, $A$ is a positive matrix. The rest of the proof
depends only on the property $A \geq 0$ and not on the details of
$A$. To complete the argument, we now have
\begin{eqnarray*}
        \lrabig{ \vth f, (\dsl-m)^{-1} f }
        &=&
        \lrabig{ \eps^* f, \ga_0 (\dsl-m)^{-1} f }
        \\
        && \hspace{-0.8in} = \
        -\int_{x_0 < 0} dx \int_{y_0 > 0} dy \,
        \lrabig{ f(-x_0,\vec x),\ [\ga_0 (\dsl_x + m)C](x-y) f(y) }
        \\
        && \hspace{-0.8in} = \
        -\int_{x_0 < 0} dx \int_{y_0 > 0} dy \int d\vec p \,
        \lraBig{ f(-x_0,\vec x),\ A \f{e^{-|x_0-y_0|\om}}{\om}\ e^{-i \vec p \cdot (\vec x - \vec y)}f(y)}
        \\
        && \hspace{-0.8in} = \
        \int d\vec p\ \absBig{ \int e^{i \vec p \cdot \vec x -x_0 \om} \lrpBig{\f{A}{\om}}^{1/2} f(x)dx}^2
        \geq 0 .
\end{eqnarray*}
\end{proof}

\begin{proof}[Proof (potential theory)]
We will now give a second proof of \er{RPDirac}, which follows the
proof of Theorem \ref{thm:RP-Dirac-Curved}. Rather than
integrating out $p_0$ in the Fourier transform, we will instead
use integration by parts to reduce the expression to a boundary
term. Note that $\{ \vth, \dsl \} = 0$, as may be proved directly,
or deduced as a special case of \er{anticom}.

Define $u = (\dsl-m)^{-1} f$ and let $u^\vth = \vth u = \ga^0
\eps^* u$. Then
\begin{eqnarray}
%    \label{ibp1}
    \< \vth f, (\dsl-m)^{-1} f\> &=&
    \int_{\Om_-} \< \vth (\dsl-m) u, u\> \ dx
    =
    -\int_{\Om_-} [\< \dsl u^\vth, u\> + m\< u^\vth, u\>] \ dx
%    \label{ibp2}
    \nn\\
    &=&
    -\int_{\Om_-} [\<\dsl u^\vth, u\> + \<u^\vth, (\dsl-m) u\> + m\< u^\vth, u\>] \ dx
    \label{ibp3}
    \\
    &=&
    -\int_{\Om_-} [\<\dsl u^\vth, u\> + \<u^\vth, \dsl u\>] \ dx
    \label{ibp4}
\end{eqnarray}
To obtain \er{ibp3}, we used that $(\dsl-m) u = f$ is zero on
$\Om_-$. \footnote{It is interesting that in \er{ibp3}, the
explicit mass term cancels out. Thus all of the $m$-dependence is
contained in $u$, which depends implicitly on $m$ through the
equation $(\dsl-m)u = f$.}

Now perform integration by parts on the first term in \er{ibp4},
moving the $\dsl$ onto the $u$. All of the non-surface terms
cancel. There is no boundary in the spatial directions, so we only
need to compute the boundary term which occurs at the $t = 0$
plane. To do this, consider
\[
    - \int_0^\infty dx_0\ \<\ga^0 \d_0 (\vth u), u\>
    =
    \int_{x_0 = 0} |u|^2 \, d^{d-1}x
    \ \ + \ \
    \lrpbig{\text{non-surface terms}}  \, .
\]
Here we used that $\vth u = \ga^0 u$ on the boundary, and
$(\ga^0)^2 = I$. Then
\[
    \lrabig{ \vth f, (\dsl-m)^{-1} f}
    =
    \int_{x_0 = 0} |u|^2 \, d^{d-1}x \geq 0
    \, .
\]
The resulting formula for the fermionic inner product is the
special case of \er{final-positivity} with $F = 1$.
\end{proof}

\section{Further Directions}

A more subtle question arises when one asks whether one can obtain
a representation of these Euclidean Green's functions as
expectations of ``classical'' Euclidean fields.  Berezin proposed
some time ago that classical fermion fields take values in a
Grassmann algebra.

Osterwalder and Schrader demonstrated that one can have Euclidean
Dirac fields. But they showed that one must double the number of
degrees of freedom; in this way they introduced a Euclidean Dirac
field $\Psi$ that is independent from (anti-commutes with) its
Dirac adjoint field $\overline\Psi$. The existence of a
representation of the Dirac propagator as an expectation of
products of Euclidean fields on curved spacetimes is, at present,
an open question. The results in this paper show that a
representation in terms of Euclidean Fermi fields is a reasonable
thing to expect.

\end{document}